\documentstyle[graphicx,times]{jaa}
\newcommand{\disp}{\displaystyle}

\begin{document}

\title[Parametric models of the periodogram]{Parametric models of the periodogram}

\author[P.\ Mohan, A.\ Mangalam \& S. Chattopadhyay]{P.\ Mohan$^{1}$, A.\ Mangalam$^{1}$ \thanks {E-mail: (prashanth, mangalam)@iiap.res.in} and S. \ Chattopadhyay$^{2}$ \thanks {Email: sudip$\_$chattopadhyay@rediffmail.com}\\ $^{1}$Indian Institute of Astrophysics, Sarjapur Road, Koramangala, Bangalore - 560034, India.\\ $^{2}$ Bengal Engineering and Science University, Shibpur, Howrah - 711103, India.}
\date{Accepted . Received ; in original form }
\pagerange{\pageref{firstpage}--\pageref{lastpage}} \pubyear{2013}
\maketitle
\label{firstpage}

\begin{abstract}
The maximum likelihood estimator is used to determine fit parameters for various parametric models of the Fourier periodogram followed by the selection of the best fit model amongst competing models using the Akaike information criteria. This analysis, when applied to light curves of active galactic nuclei can be used to infer the presence of quasi-periodicity and break or knee frequencies. The extracted information can be used to place constraints on the mass, spin and other properties of the putative central black hole and the region surrounding it through theoretical models involving disk and jet physics. 
\end{abstract}

\section{Introduction \& periodogram fit models}
Broadband flux variability arises from disk and jet based phenomena. The periodogram (eg. van der Klis 1989) of light curves (LCs) of active galactic nuclei implicitly reflects this. Parametric models of the binned periodogram approximate the true expected shape, the power spectral density (PSD) which is ideally normally distributed making it useful to work with. A statistically appropriate model aids in theoretical modeling of variability.

Commonly used parametric models include: power law model (eg. Papadakis \& Lawrence 1993), $I(f)=N f^{-\alpha}$ where $\alpha$ is the red-noise slope in the range of -1 and -2.5; broken power law model (eg. Uttley et al. 2002), $I(f)=N (f/f_{Brk})^{-\alpha_{hi}} , f>f_{Brk}$ and $I(f)=N (f/f_{Brk})^{-\alpha_{low}} , f<f_{Brk}$ where $f_{Brk}$ is the break frequency, $\alpha_{hi}$ is the slope of the high frequency region and $\alpha_{low}$ is the slope of the low frequency region; bending power law model (eg. Uttley et al. 2002), $I(f)=N (1+(f/f_{Knee})^2)^{-\alpha/2}$ where $f_{Knee}$ is the knee frequency, $\alpha$ is the slope in the high frequency region above the knee frequency; power law with a Lorentzian QPO model (eg. Nowak 2000), $I(f)=N f^{-\alpha}$+$\disp{\frac{R^2 Q f^2_0/\pi}{f^2_0+Q^2 (f-f_0)^2}}$ where $\alpha$ is red-noise slope and the second term is a Lorentzian function with amplitude $R$, central frequency $f_0$ and quality factor $Q = f_0/\Delta f$ where $\Delta f$ is the frequency spread in the bin hosting the central frequency.

\section{Fit procedure, model selection \& significance testing}
The likelihood and log-likelihood functions are:
\small
\begin{eqnarray}
L(\theta_k) &=& \prod^{(n-1)}_{j=1} \frac{1}{I(f_j,\theta_k)} e^{-P(f_j)/I(f_j,\theta_k)}\\ \nonumber
 S(\theta_k) &=& 2 \sum^{n-1}_{j=1}(\ln(I(f_j,\theta_k))+P(f_j)/I(f_j,\theta_k))
\end{eqnarray}
\normalsize
$L(\theta_k)$ is the likelihood function, $\theta_k$ are the parameters of the parametric models, $I(f_j,\theta_k)$, to be estimated and $P(f_j)$ is the data periodogram. Determining $\theta_k$ which minimize $S$ yields the maximum likelihood values. Confidence levels are determined in a similar manner as that described in Press et al. (2007) for the $\Delta \chi^2$ method. Model selection is carried out using the Akaike information criteria (AIC) and the relative likelihood (RL). The AIC measures the information lost when a model is fit to the data. The model with least information loss (least AIC) is assumed to be the null model. The likelihood of other models describing the data as well as the null is determined using the the relative likelihood (eg. Burnham \& Anderson 2004). The AIC and RL are defined as:
\small
\begin{eqnarray}
AIC&=&S(\theta_k)+2 p_k \\ \nonumber
\Delta_i&=&AIC_{\mathrm{min (model \ i)}}-AIC_{\mathrm{min(null)}} \\ \nonumber
L(\mathrm{model \ i|data})&=&e^{-\Delta_i/2} \\ \nonumber
RL &=& 1/L(\mathrm{model \ i|data}) 
\end{eqnarray}
\normalsize
Where $p_k$ is a penalty term = number of parameters $\theta_k$ used in the model, $L(\mathrm{model \ i|data})$: likelihood of model $i$ given the data, $RL$:likelihood ratio of model $i$ relative to that of the null. Models with $\Delta_i\leq 2$ are close to the best fit, those with $4 \leq \Delta_i \leq 7$ are considerably less supported and those with $\Delta_i > 10$ and $RL > 150$ cannot be supported (Burnham \& Anderson 2004) in which case the null remains the best fit. On subtraction of the PSD from it, the periodogram ideally comprises of $\chi^2_2$ distributed white noise (eg. Chatfield 2009). The residue from the best fit to the data is tested against the $\chi^2_2$ distribution using a goodness of fit test. A 99\% significance level (based on its cumulative distribution function) then reveals the outlying QPO.

\section{Application, results \& conclusions}
\begin{figure}
\centerline{\includegraphics[scale=0.35]{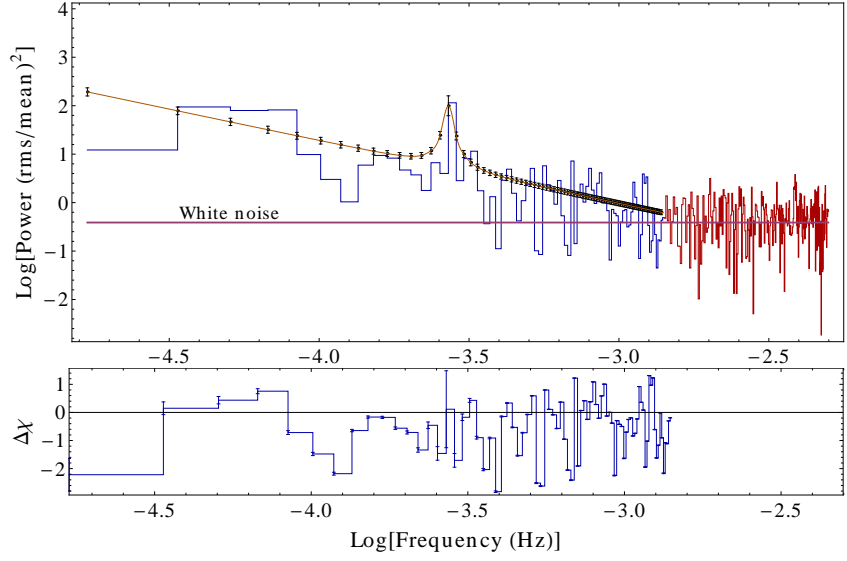}}
\caption{Binned periodogram: fit portion is in blue and white noise region is in red. The best fit is the power law with Lorentzian QPO model and the residue $\Delta \chi=$(data-model)/$\sigma$ is shown below it.}
\label{fig1}
\end{figure}

The XMM Newton X-ray light curve (0.3 keV - 10 keV) of REJ 1034+396 which revealed a QPO of $\sim$ 3733 s (Gierli{\'n}ski et al. 2008) is analyzed using the periodogram (296 bins) with the above fit models (results in Table \ref{tab1}). The power law with QPO model (Fig. \ref{fig1}) is the best fit model with an AIC = 149.8 and a significance of the QPO $>$ 99.94 \%. Parameter estimation with MLE and model selection with AIC is computationally efficient when compared to Monte-Carlo simulations based procedures (Uttley et al. 2002). Any other parametric model can be easily incorporated into this procedure. If a statistically significant QPO is detected, a lower limit to the black hole mass and constraints on its spin can be placed assuming that the QPO is from an orbital signature. Theoretical models considering general relativistic effects and emission region structure can be used to simulate PSDs which can then be compared with observations to yield constraints on physical parameters.

\bibliography{}

\begin{table}
\centering
\resizebox{10cm}{5cm}{
\begin{tabular}{|l|l|l|}
\hline
Fit model         & Properties & Results\\
\hline
 & & \\
Power law         & (AIC, RL) & (170.1, 25.3 $\times$ 10$^3$) \\
Fit parameters    & Power law slope ($\alpha$) & (-1.33 $\pm$ 0.013) \\
                  & Significance (\%) &  $>$99.99\\   
\hline
 & & \\
Broken power law  & (AIC, RL) & (172.1, 68.1 $\times$ 10$^3$) \\
Fit parameters    & Break frequency $f_{Brk}$ (Hz) & 0.00032 $\pm$ 0.000074\\
                  & High frequency region slope $\alpha_{hi}$ & -1.6 $\pm$ 0.42\\
                  & Low frequency region slope $\alpha_{low}$ & -1 $\pm$ 0.33\\
                  & Significance (\%) &  $>$99.99\\   
\hline
 & & \\
Bending power law & (AIC, RL) & (198.4, 3.5 $\times$ 10$^{10}$) \\
Fit parameters    & Knee frequency $f_{Knee}$ (Hz) & 0.00084 $\pm$ 0.00018\\
                  & High frequency region slope $\alpha$ & -3.9 $\pm$ 0.74\\
                  & Significance (\%) &  $>$99.99\\   
\hline
 & & \\
Power law \& QPO  & (AIC, RL) & (149.8, 1)\\
Fit parameters    & Central frequency $f_0$ (Hz) & 0.000269\\
                  & Amplitude R & 0.05 $\pm$ 0.014\\
                  & Quality factor Q & 32 $\pm$ 6.5\\
                  & Significance (\%) &  $>$99.94\\                                     
\hline
\end{tabular}}
\caption{Results of the periodogram analysis with fit models.}
\label{tab1}
\end{table}

\end{document}